\documentclass[12pt]{article}
\usepackage{amsfonts}
\usepackage{amssymb}
\usepackage{graphics,psboxit,amsmath}

\def\hybrid{\topmargin 0pt    \oddsidemargin 0pt
        \headheight 0pt \headsep 0pt
        \textwidth 6.35in       % BS paper
        \textheight 9.10in       % BS paper
        \marginparwidth .875in
        \parskip 5pt plus 1pt   \jot = 1.5ex}

\hybrid

\def\baselinestretch{1.2}

\catcode`\@=11

\def\marginnote#1{}
\newcount\hour
\newcount\minute
\newtoks\amorpm
\hour=\time\divide\hour by60
\minute=\time{\multiply\hour by60 \global\advance\minute by-\hour}
\edef\standardtime{{\ifnum\hour<12 \global\amorpm={am}%
        \else\global\amorpm={pm}\advance\hour by-12 \fi
        \ifnum\hour=0 \hour=12 \fi
        \number\hour:\ifnum\minute<10 0\fi\number\minute\the\amorpm}}
\edef\militarytime{\number\hour:\ifnum\minute<10 0\fi\number\minute}
\def\draftlabel#1{{\@bsphack\if@filesw {\let\thepage\relax
   \xdef\@gtempa{\write\@auxout{\string
      \newlabel{#1}{{\@currentlabel}{\thepage}}}}}\@gtempa
   \if@nobreak \ifvmode\nobreak\fi\fi\fi\@esphack}
        \gdef\@eqnlabel{#1}}
\def\@eqnlabel{}
\def\@vacuum{}
\def\draftmarginnote#1{\marginpar{\raggedright\scriptsize\tt#1}}

\def\draft{\oddsidemargin -.5truein
        \def\@oddfoot{\sl preliminary draft \hfil
        \rm\thepage\hfil\sl\today\quad\militarytime}
        \let\@evenfoot\@oddfoot \overfullrule 3pt
        \let\label=\draftlabel
        \let\marginnote=\draftmarginnote
   \def\@eqnnum{(\theequation)\rlap{\kern\marginparsep\tt\@eqnlabel}%
\global\let\@eqnlabel\@vacuum}  }

\def\preprint{\twocolumn\sloppy\flushbottom\parindent 2em
        \leftmargini 2em\leftmarginv .5em\leftmarginvi .5em
        \oddsidemargin -.5in    \evensidemargin -.5in
        \columnsep .4in \footheight 0pt
        \textwidth 10.in        \topmargin  -.4in
        \headheight 12pt \topskip .4in
        \textheight 6.9in \footskip 0pt
        \def\@oddhead{\thepage\hfil\addtocounter{page}{1}\thepage}
        \let\@evenhead\@oddhead \def\@oddfoot{} \def\@evenfoot{} }

\def\numberbysection{\@addtoreset{equation}{section}
        \def\theequation{\thesection.\arabic{equation}}}

\def\underline#1{\relax\ifmmode\@@underline#1\else
        $\@@underline{\hbox{#1}}$\relax\fi}

\def\titlepage{\@restonecolfalse\if@twocolumn\@restonecoltrue\onecolumn
     \else \newpage \fi \thispagestyle{empty}\c@page\z@
        \def\thefootnote{\fnsymbol{footnote}} }

\def\endtitlepage{\if@restonecol\twocolumn \else \newpage \fi
        \def\thefootnote{\arabic{footnote}}
        \setcounter{footnote}{0}}  %\c@footnote\z@ }

\catcode`@=12
\relax

\def\figcap{\section*{Figure Captions\markboth
        {FIGURECAPTIONS}{FIGURECAPTIONS}}\list
        {Figure \arabic{enumi}:\hfill}{\settowidth\labelwidth{Figure
999:}
        \leftmargin\labelwidth
        \advance\leftmargin\labelsep\usecounter{enumi}}}
 \relax
\def\tablecap{\section*{Table Captions\markboth
        {TABLECAPTIONS}{TABLECAPTIONS}}\list
        {Table \arabic{enumi}:\hfill}{\settowidth\labelwidth{Table
999:}
        \leftmargin\labelwidth
        \advance\leftmargin\labelsep\usecounter{enumi}}}
 \relax
\def\reflist{\section*{References\markboth
        {REFLIST}{REFLIST}}\list
        {[\arabic{enumi}]\hfill}{\settowidth\labelwidth{[999]}
        \leftmargin\labelwidth
        \advance\leftmargin\labelsep\usecounter{enumi}}}
 \relax
\makeatletter
\newcounter{pubctr}
\def\publist{\@ifnextchar[{\@publist}{\@@publist}}
\def\@publist[#1]{\list
        {[\arabic{pubctr}]\hfill}{\settowidth\labelwidth{[999]}
        \leftmargin\labelwidth
        \advance\leftmargin\labelsep
        \@nmbrlisttrue\def\@listctr{pubctr}
        \setcounter{pubctr}{#1}\addtocounter{pubctr}{-1}}}
\def\@@publist{\list
        {[\arabic{pubctr}]\hfill}{\settowidth\labelwidth{[999]}
        \leftmargin\labelwidth
        \advance\leftmargin\labelsep
        \@nmbrlisttrue\def\@listctr{pubctr}}}
 \relax
\makeatother
\newskip\humongous \humongous=0pt plus 1000pt minus 1000pt

\newif\ifdtup

\relax

\def\be{\begin{equation}}
\def\ee{\end{equation}}
\def\ba{\begin{eqnarray}}
\def\ea{\end{eqnarray}}

\def\no{\noindent}

\def\IR{\relax{\rm I\kern-.18em R}}
\def\II{\relax{\rm 1\kern-.35em1}}

\renewcommand{\theequation}{\thesection.\arabic{equation}}
\csname @addtoreset\endcsname{equation}{section}

\def\IR{\relax{\rm I\kern-.18em R}}
\def\inv{^{\raise.15ex\hbox{${\scriptscriptstyle -}$}\kern-.05em 1}}

\begin{document}

\begin{titlepage}
\begin{center}

%\hfill  arXiv:yymm.nnnn [hep-th]\\

\vskip 1in

{\LARGE Three-point correlators for giant magnons}
\vskip 0.4in

{\bf Rafael Hern\'andez}
 
\vskip 0.1in

Departamento de F\'{\i}sica Te\'orica I\\
Universidad Complutense de Madrid\\
$28040$ Madrid, Spain\\
{\footnotesize{\tt rafael.hernandez@fis.ucm.es}}

\end{center}

\vskip .4in

\centerline{\bf Abstract}
\vskip .1in
\no
Three-point correlation functions in the strong-coupling regime of the AdS/CFT correspondence can be analyzed 
within a semiclassical approximation when two of the vertex operators correspond to heavy string states having large 
quantum numbers while the third vertex corresponds to a light state with fixed charges.
We consider the case where the heavy string states are chosen to be giant magnon solitons with either a single or two 
different angular momenta, for various different choices of light string states.

\noindent

\vskip .4in
\noindent

\end{titlepage}
\vfill
\eject

\def\baselinestretch{1.2}

%%%%%%%%%%%%%%%%%%%%%%%%%%%%%%%%%%%%%%%%%%%%%%%%%%%%%%%%%%%%%%%%%%%%%%%%%%%%%%%%%%%%%%

\baselineskip 20pt

%%%%%%%%%%%%%%%%%%%%%%%%%%%%%%%%%%%%%%%%%%%%%%%%%%%%%%%%%%%%%%%%%%%%%%%%%%%%%%%%%%%%%%
%%%%%%%%%%%%%%%%%%%%%%%%%%%%%%%%%%%%%%%%%%%%%%%%%%%%%%%%%%%%%%%%%%%%%%%%%%%%%%%%%%%%%%

\section{Introduction}

A conformal field theory is entirely determined once the complete spectrum of two and three-point correlation functions is solved for any 
value of the coupling constant. Higher order correlators can then be obtained from these two lower ones. The appearance of integrable 
structures in the AdS/CFT correspondence provided an impressive analysis of two-point correlation functions and the spectrum of anomalous 
dimensions in four-dimensional planar Yang-Mills with ${\cal N}~=~4$ supersymmetry both in the weak and strong-coupling regimes (see for instance~\cite{reviewint} 
for a comprehensive review) . There is however no equivalent understanding on the general structure of three-point correlation functions. In the 
weak-coupling limit three-point correlators have been evaluated perturbatively~\cite{perturbative1} or using integrability-inspired techniques~\cite{threepointint,Vieira}.
In the strong-coupling limit three-point functions for chiral operators have been computed within the supergravity regime of the 
correspondence~\cite{correlationsugra1}. But it has been only recently that a general analysis has started for primary operators dual to massive string states. 
In the AdS/CFT correspondence, the strong-coupling limit of correlation functions for single-trace gauge invariant operators can be 
found by inserting closed string vertex operators in the path integral for the string partition function. These vertex operators scale exponentially 
with both the energy and the quantum conserved charges for the corresponding string states. Therefore, when charges are of the order 
of the string tension a saddle point approximation can be used in order to evaluate the string path integral, which will be dominated by a semiclassical 
string trajectory. We can thus employ a semiclassical approximation in order to analyze the case of correlation functions for non-protected operators 
with large quantum charges. 
  
The semiclassical approach to the evaluation of two-point correlation functions was explored in references \cite{GKP}-\cite{BT}. The analysis of three-point 
correlators where two of the vertex operators are complex conjugate {\em heavy} string states with large conserved charges while the third one is a {\em light} state 
with fixed charges has been developed in a recent series of papers \cite{Zarembo}-\cite{Arnaudov}.~\footnote{A semiclassical treatment of four-point 
correlation functions with two heavy and two light vertex operators has also been considered in \cite{BTfourpoint,Arnaudov}.} 
In the case where the light vertex is a massless mode corresponding to a protected chiral state it was shown in \cite{Zarembo,Costa} that indeed 
the leading order contribution to the correlator in the large string tension limit is coming from the semiclassical trajectory, which amounts to evaluating 
the light vertex operator on the classical string configuration of the two heavy vertices. The extension to the case where the light vertex operator 
is a massive string mode, dual to general non-protected states, was later on considered in \cite{RT}. 
  
The leading contribution in the saddle point approximation to the correlation function of two complex conjugate heavy vertex operators and one light vertex is thus coming 
from the classical string configuration of the operators with large quantum charges. Therefore the contribution from the light vertex can be neglected  and the three-point 
correlator is governed by the classical solution saturating the two-point function of the heavy vertices. In order to find the three-point function $\langle V_{H_1}(x_1) V_{H_2}(x_2) V_{L}(x_3) \rangle$ it 
suffices then to evaluate the light vertex operator on the classical configuration,
\be
\langle V_{H_1}(x_1) V_{H_2}(x_2) V_{L}(x_3) \rangle = V_L(x_3)_{\hbox{\tiny{classical}}} \ .
\ee
In a conformal field theory the dependence on the location of the vertex operators in a three-point function is completely determined up to some overall 
coefficients $C_{123}$, which are the structure constants in the operator product expansion. In order to find the value of these coefficients the positions of the vertex 
operators can be conveniently chosen \cite{BT,RT}. Taking $|x_1|=|x_2|=1$ and $x_3=0$,  the correlator reduces to
\be
\langle V_{H_1}(x_1) V_{H_2}(x_2) V_{L}(0) \rangle = \frac {C_{123}} {|x_1-x_2|^{2\Delta_{H_1}}} \ ,
\ee
because the conformal weights for the heavy operators are much larger than that of the light operator. The normalized structure constant 
${\cal C}_3 \equiv C_{123}/C_{12}$ can then obtained from
\be
{\cal C}_3 = c_{\Delta}  V_L(0)_{\hbox{\tiny{classical}}} \ ,
\ee
where $c_{\Delta}$ is the normalization constant of the corresponding light vertex operator. In this note we will employ this proposal to explore 
the case where the classical states associated with the the heavy vertices in the three-point correlator are giant magnon solitons with a single or two 
different angular momenta in $S^5$. 
The correlation function of two single-charge giant magnons and the lagrangian operator has been considered before in \cite{Costa}.
The purpose of this article is to extend the analysis to more general three-point functions.

The remaining part of this note is organized as follows. In section 2 we briefly review some general features of the giant magnon solutions. 
Then in section 3 we will compute the normalized coefficients in the three-point functions for either the single or the two-charge giant magnon solitons, and 
various different choices of light vertex operators. In section 4 we conclude with some general remarks and a discussion on several open problems.

%%%%%%%%%%%%%%%%%%%%%%%%%%%%%%%%%%%%%%%%%%%%%%%%%%%%%%%%%%%%%%%%%%%%%%%%%%%%%%%%%%%%%%
%%%%%%%%%%%%%%%%%%%%%%%%%%%%%%%%%%%%%%%%%%%%%%%%%%%%%%%%%%%%%%%%%%%%%%%%%%%%%%%%%%%%%%

\section{Giant magnons}

In this section we will briefly present the giant magnon solitons for the string sigma model with either a single or two different angular 
momenta in $S^5$. To describe the solutions, it will be convenient to parameterize the embedding coordinates $Y_M$ and $X_K$ for the 
ten-dimensional $AdS_{5} \times S^5$ background in terms of the corresponding global angles,
\ba
Y_1 + i Y_2 \! \! \! & = & \! \! \! \sinh \rho \sin \gamma e^{i\phi_1} \  , \: \: Y_3 + i Y_4 = \sinh \rho \cos \gamma e^{i\phi_2} \  , 
\: \: Y_5 + i Y_0 = \cosh \rho \, e^{i t} \  ,  \\
X_1 + i X_2  \! \! \! & = & \! \! \! \cos \theta \cos \psi e^{i\tilde{\varphi}} \  , \: \: X_3 + i X_4 = \cos \theta \sin \psi e^{i\bar{\varphi}} \  , 
\: \: X_5 + i X_6 = \sin \theta e^{i\varphi} \  . 
\ea
The embedding coordinates of $AdS_5$ are related to the Poincar\'e coordinates through
\be
Y_m = \frac {x_m}{z} \ , \quad Y_4 = \frac {1}{2z} \big( -1 + z^2 + x^m x_m \big) \ , \quad 
Y_5 = \frac {1}{2z} \big( 1 + z^2 + x^m x_m \big) \ ,
\ee
where $x^m x_m = -x_0^2 + x_i x_i$, with $m=0,1,2,3$ and $i=1,2,3$.
Euclidean continuation of the time-like directions to
\be
t_e = i t \ , \quad Y_{0e} = i Y_0 \ , \quad x_{0e} = i x_0 \ ,
\ee
will allow the classical trajectories to approach the boundary $z=0$ when $\tau_e=\pm \infty$. The giant magnon with a single angular momentum is a localized 
classical soliton propagating on an infinite string moving in $\mathbb{R} \times S^2$ \cite{HM}. After euclidean rotation it is described by
\ba
z & \!\!\! = \!\!\! &  \hbox{sech} (\kappa \tau_{e}) \ , \quad x_{0e} = \tanh (\kappa \tau_{e}) \ , \quad x_i = 0 \ , \label{Poincarecoordinates} \\
\cos \theta & \!\!\! = \!\!\! & \sin \frac {p}{2} \hbox{sech}( u_{e} ) \ , \quad \tan (\varphi + i \tau_{e}) = \tan \frac {p}{2} \tanh ( u_{e} )  \ , 
\label{giantmagnon}
\ea
with $p$ the momentum of the magnon and 
\be
u_{e} = \big( \sigma + i \tau_{e} \cos \frac {p}{2} \big)  \csc \frac {p}{2} \ . 
\label{u}
\ee
The Virasoro constraint requires $\kappa^2 = 1$, which also follows from the marginality condition for conformal invariance on the semiclassical 
two-point correlation function of the corresponding physical vertex operators. Both the energy $E$ and the angular momentum $J$ for the giant magnon 
are infinite, but the difference is kept finite,
\be
E - J = \frac {\sqrt{\lambda}}{\pi} \left| \sin \frac {p}{2} \right| \ ,
\label{EJ}
\ee
which is the strong-coupling limit of the dispersion relation for elementary magnon excitations in four-dimensional planar ${\cal N}=4$ Yang-Mills \cite{dispersionrelation}.
  
The case of a giant magnon soliton with two different angular momenta corresponds to a classical string solution rotating in $\mathbb{R} \times S^3$ 
\cite{Dorey}-\cite{Bobev}. We will present the solution following closely notation and conventions in reference \cite{dressing}. After euclidean continuation 
the giant magnon with two charges is described by the point-like $AdS_5$ geodesic (\ref{Poincarecoordinates}), together with 
\ba
\cos \theta & \!\!\! = \!\!\! & \sin \frac {p}{2} \hbox{sech}( v_{e} ) \ , \quad \tan (\varphi + i \tau_{e}) = \tan \frac {p}{2} \tanh ( v_{e} )  \ , \nonumber \\
\bar{\varphi} & \!\!\! = \!\!\! & - ( \sigma \sinh \alpha + i \tau_e \cosh \alpha) \sin \beta \ ,
\label{twogiantmagnon}
\ea
where $p$ is again the momentum of the magnon, and now
\ba
v_e = \big( \sigma \cosh \alpha + i \tau_e \sinh \alpha \big) \cos \beta \ , 
\ea
with the parameters $\alpha$ and $\beta$ given by
\be
\tanh \alpha = \frac {2r}{1+r^2} \cos \frac {p}{2} \ , \quad \cot \beta = \frac {2r}{1-r^2} \sin \frac {p}{2} \ .
\label{alphabeta}
\ee
The conserved finite charges carried by the two-charge giant magnon soliton are
\ba
E - J \!\! & = & \!\! \frac {\sqrt{\lambda}}{\pi} \frac {1+r^2}{2r} \left| \sin \frac {p}{2} \right| \ , \\ \label{EJtwo}
\bar{J} \!\!  & = & \!\! \frac {\sqrt{\lambda}}{\pi} \frac {1-r^2}{2r} \left| \sin \frac {p}{2} \right| \ . \label{barJ}
\ea
Eliminating the parameter $r$ in these expressions we get 
\be
E -J = \sqrt{\bar{J}^2 + \frac {\lambda}{\pi^2} \sin^2 \frac {p}{2}} \ ,
\label{bounddispersion}
\ee
which is the dispersion relation for a bound state of $\bar{J}$ giant magnons \cite{Dorey}. In the limit where the parameter $r \rightarrow 1$ 
the second angular momentum $\bar{J}$ vanishes, and the two-charge soliton reduces to the elementary giant magnon with a single angular momentum. 

%%%%%%%%%%%%%%%%%%%%%%%%%%%%%%%%%%%%%%%%%%%%%%%%%%%%%%%%%%%%%%%%%%%%%%%%%%%%%%%%%%%%%%
%%%%%%%%%%%%%%%%%%%%%%%%%%%%%%%%%%%%%%%%%%%%%%%%%%%%%%%%%%%%%%%%%%%%%%%%%%%%%%%%%%%%%%

\section{Three-point correlation functions}

In this section we will find the leading contribution in the large string tension limit to three-point correlation functions where the complex conjugate heavy vertices 
correspond to the giant magnon solitons described in the previous section, and the light vertex is an operator carrying quantum conserved charges of the order 
of unity. In order to find the normalized structure constants we will follow closely the general proposal in reference \cite{RT} for various different choices of light 
string states.

\subsection{Dilaton operator} 
  
We will first analyze the case where the light vertex is taken to be the massless dilaton operator. In the large string tension expansion the leading contribution 
to the massless dilaton vertex is just bosonic,
\be
V^{\hbox{\tiny{(dilaton)}}} = (Y_+)^{-\Delta_d} \, (X_x)^j 
\big[ z^{-2} (\partial x_m \bar{\partial}x^m + \partial z \bar{\partial} z ) + \partial X_K \bar{\partial} X_K \big] \ ,
\label{Vdilaton}
\ee
where here and along this note we have defined $Y_+ \equiv Y_4 + Y_5$ and $X_x \equiv X_1 + i X_2$, and the derivatives are $\partial \equiv \partial_+$ 
and $\bar{\partial} \equiv \partial_-$. To leading order the scaling dimension in the strong-coupling regime is $\Delta_d=4+j$, 
where $j$ denotes the Kaluza-Klein momentum of the dilaton. The corresponding dual gauge invariant operator is $\hbox{Tr}(F_{\mu \nu}^2 Z^j + \cdots )$.
  
We will first consider the correlator where the heavy vertex operators are giant magnons solitons with a single angular momentum.
As explained above, the three-point correlation function is dominated by the light vertex operator evaluated on the classical trajectory provided by the giant magnon 
solutions. Therefore in order to find the contribution from the magnon to the dilaton vertex operator we use relations (\ref{Poincarecoordinates}) and (\ref{giantmagnon}) 
to compute
\be
z^{-2} ( \partial x_m \bar{\partial}x^m + \partial z \bar{\partial} z ) = \kappa^2 \ , \quad
\partial X_K \bar{\partial} X_K = 2 \, \hbox{sech}^2 (u_{e}) - 1 \ .
\ee
The normalized coefficient in the three-point correlator is then obtained from
\be
{\cal C}_{3}^{\hbox{\tiny{(dilaton)}}} = c_{\Delta}^{\hbox{\tiny{(dilaton)}}} \int_{-\infty}^{\infty} d \tau_e \int_{-\infty}^{\infty} d \sigma \, 
(\cos \theta)^j (\hbox{sech} (\kappa \tau_e) )^{\Delta_d} \big( \kappa^2 + 2 \, \hbox{sech}^2 (u_{e}) - 1 \big) \ ,
\label{Cdilatonintegral}
\ee
where $c_{\Delta}^{\hbox{\tiny{(dilaton)}}}$ is the normalization constant of the dilaton vertex operator \cite{Berenstein,Zarembo},
\be
c_{\Delta}^{\hbox{\tiny{(dilaton)}}} = \frac {2^{-j} \sqrt{\lambda}}{128 \pi N} \sqrt{(j+1)(j+2)(j+3)} \ .
\ee
The integrations over $\sigma$ and $\tau$ in expression (\ref{Cdilatonintegral}) factorize, and both integrals turn to be essentially the same. If we define
\be
I{(\Delta, j)} = \int_{-\infty}^{\infty} d \tau_e \left( \hbox{sech} (\kappa \tau_{e})\right)^{\Delta} \int_{-\infty}^{\infty} d \sigma \left( \hbox{sech} (u_{e}) \right)^j \ ,
\label{I}
\ee
the coefficient in the correlator can be written as
\be
{\cal C}_{3}^{\hbox{\tiny{(dilaton)}}} = c_{\Delta}^{\hbox{\tiny{(dilaton)}}} \, 
\big( (\kappa^2 - 1) I{(\Delta_d,j)} + 2 I{(\Delta_d,j+2)} \big) \left( \sin \frac {p}{2} \right)^j \ .
\ee
The integrals can be easily computed using that
\be
\int  d x \left( \hbox{sech} (ax) \right)^b  = \frac {1}{a} \tanh(ax) \,
{}_2F_1 \left( \frac {1}{2}, 1 - \frac {b}{2}, \frac {3}{2} ; \tanh^2(ax) \right) \ .
\ee
Therefore the integration over $\sigma$ becomes
\be
\int_{-\infty}^{\infty} d \sigma \left( \hbox{sech} (u_{e}) \right)^j  = \sqrt{\pi} \frac {\Gamma(j/2)}{\Gamma(1/2+j/2)} \sin \frac {p}{2} \ ,
\label{Isigma}
\ee
while the integral over $\tau_e$ is
\be
\int_{-\infty}^{\infty} d \tau_e \left( \hbox{sech} (\kappa \tau_{e}) \right)^{\Delta_d} = \frac {\sqrt{\pi}}{\kappa} \frac {\Gamma(\Delta_d/2)}{\Gamma(1/2+\Delta_d/2)} \ .
\label{Itau}
\ee
The coefficient in the three-point correlation function with two single-charge giant magnon vertices and a light dilaton vertex is then
\be
{\cal C}_{3}^{\hbox{\tiny{(dilaton)}}} = 2 \pi c_{\Delta}^{\hbox{\tiny{(dilaton)}}} 
\frac {\Gamma \left( 1 + j/2 \right) \Gamma \left( 2 + j/2 \right)}{\Gamma \left( 3/2 + j/2 \right) \Gamma \left( 5/2 + j/2 \right)} \left( \sin \frac {p}{2} \right)^{j+1} \ ,
\label{Cdilaton}
\ee
where we have made use of the marginality condition of the semiclassical vertex operators. 
When $j=0$ the coupling is just to the lagrangian, which is the correlator analyzed in \cite{Costa}. 
Expression (\ref{Cdilaton}) reduces then to
\be
{\cal C}_{3, j=0}^{\hbox{\tiny{(dilaton)}}} = 
\frac {\sqrt{6}}{24 \pi} \frac {\sqrt{\lambda}}{N} \sin \frac {p}{2} \ .
\ee
Recalling now the dispersion relation (\ref{EJ}) we recover the observation in reference \cite{Costa} that the coefficient in the three-point correlation function 
when the light vertex is the lagrangian operator is proportional to the derivative with respect to $\lambda$ of the anomalous dimension for the 
giant magnon. This relation seems to be quite a general result, as argued from a thermodynamical point of view in \cite{RT2}, and noticed for several 
different choices of heavy string states in \cite{RT}-\cite{Ryang}. 
  
Let us extend now the above analysis to the case where the heavy vertex operators are giant magnon solitons with two different angular momenta. 
Using solution (\ref{twogiantmagnon}) we get
\be
\partial X_K \bar{\partial} X_K = 2 \cos^2 \beta \, \hbox{sech}^2 \big( v_{e}) - 1 \ ,
\ee
with $\beta$ as given in (\ref{alphabeta}). Using now the integral $I{(\Delta, j)}$ defined before the coefficient in the three-point correlator 
in the case of general $\Delta_d$ can be written as
\be
{\cal C}_{3}^{\hbox{\tiny{(dilaton)}}} = c_{\Delta}^{\hbox{\tiny{(dilaton)}}} \, 
\big( (\kappa^2 - 1) I{(\Delta_d,j)} + 2 \cos^2 \beta \, I{(\Delta_d,j+2)} \big) \left( \sin \frac {p}{2} \right)^j \ .
\ee
Evaluating the integral and taking into account the conformal constraint we get
\be
{\cal C}_{3}^{\hbox{\tiny{(dilaton)}}} = 2 \pi  c_{\Delta}^{\hbox{\tiny{(dilaton)}}} \frac {2r}{1+r^2}
\frac {\Gamma \left( 1 + j/2 \right) \Gamma \left( 2 + j/2 \right)}{\Gamma \left( 3/2 + j/2 \right) \Gamma \left( 5/2 + j/2 \right)} 
\left( \sin \frac {p}{2} \right)^{j+1} \ .
\label{Cdilatontwomagnon}
\ee
Recalling now relations (\ref{EJtwo}) and (\ref{bounddispersion}), in the case when $j=0$ the coefficient (\ref{Cdilatontwomagnon}) can be written as
\be
{\cal C}_{3, j=0}^{\hbox{\tiny{(dilaton)}}} = \frac {\sqrt{6}}{24 \pi^2} \frac {\lambda}{N} 
\frac {\sin^2 \frac {p}{2}}{\sqrt{\bar{J}^2 + \frac {\lambda}{\pi^2} \sin^2 \frac {p}{2}}} \ , 
\ee
which extends to the case of giant magnon solitons with two different angular momenta the above observation that the structure constant of the three-point 
correlation function for two single-charge giant magnon heavy states coupled to the lagrangian is proportional to the $\lambda$-derivative 
of the corresponding dispersion relation.

%%%%%%%%%%%%%%%%%%%%%%%%%%%%%%%%%%%%%%%%%%%%%%%%%%%%%%%%%%%%%%%%%%%%%%%%%%%%%%%%%%%%%%
%%%%%%%%%%%%%%%%%%%%%%%%%%%%%%%%%%%%%%%%%%%%%%%%%%%%%%%%%%%%%%%%%%%%%%%%%%%%%%%%%%%%%%

\subsection{Primary scalar operator}

We will now choose the light vertex to be the superconformal primary scalar operator. The bosonic part of the primary scalar vertex is \cite{Berenstein,Zarembo,RT}
\be
V^{\hbox{\tiny{(primary)}}} = (Y_+)^{-\Delta_p} (X_x)^j 
\big[ z^{-2} (\partial x_m \bar{\partial}x^m - \partial z \bar{\partial} z ) - \partial X_K \bar{\partial} X_K \big] \ ,
\label{primary}
\ee
where the scaling dimension is now $\Delta_p=j$. The corresponding operator in the dual gauge theory is the BMN operator $\hbox{Tr}Z^j$. 
When the classical trajectories from the heavy vertex operators approach BMN geodesics the correlation function should reproduce the correlator of 
three chiral primary operators. However it was noticed in reference \cite{Zarembo} that an additional anomalous term arises after the BMN-limit of the 
heavy spinning string states. In \cite{Russo} it was shown that the ambiguity implied by the anomalous rescaling of the correlation function in the large spin limit 
is a generic feature of string states with a point-like BMN limit, and can be removed through a convenient choice of normalization constant of the light chiral 
primary operator. In our analysis below we will show that in the case of giant magnon solitons there seems to be no room for ambiguities and a different choice 
of normalization for the light vertex. 

We will first consider the case where the heavy vertices are giant magnons solitons with a single angular momentum. Using (\ref{Poincarecoordinates}) we get
\be
z^{-2} (\partial x_m \bar{\partial}x^m - \partial z \bar{\partial} z ) = \kappa^2 (2 \, \hbox{sech}^2 (\kappa \tau_e) - 1) \ ,
\ee
and thus the light vertex operator becomes
\be
V^{\hbox{\tiny{(primary)}}} = 
(\cos \theta)^j (\hbox{sech} (\kappa \tau_e) )^{\Delta_p} \big[ \kappa^2 (2 \, \hbox{sech}^2 (\kappa \tau_e) - 1) - (2 \, \hbox{sech}^2 (u_e) - 1) \big] \ .
\ee
The normalized coefficient in the three-point correlator can then be written as
 \be
{\cal C}_3^{\hbox{\tiny{(primary)}}} = c_{\Delta}^{\hbox{\tiny{(primary)}}}  (\kappa^2 - 1) 
\big(2 I{(\Delta_p,j+2)} - I{(\Delta_p,j)} \big) \left( \sin \frac {p}{2} \right)^j \ ,
\ee
where we have used that the integrals, as defined in (\ref{I}) with $\Delta_p$ the scaling dimension for the chiral primary operator, now satisfy $I(\Delta_p+2,j)=I(\Delta_p,j+2)$. 
Performing the integrations we end up with
\be
{\cal C}_3^{\hbox{\tiny{(primary)}}} = \pi c_{\Delta}^{\hbox{\tiny{(primary)}}} 
\frac {(\kappa^2 - 1)}{\kappa} \frac {(j-1) \Gamma(j/2)^2}{(j+1)\Gamma(1/2+j/2)^2} \left( \sin \frac {p}{2} \right)^{j+1} \ .
\ee
When we take into account the conformal condition we find that the three-point function vanishes identically, so that there is no coupling 
between one primary scalar light operator and two single-charge giant magnon solitons. Note that as the conformal constraint comes as a global factor there 
is no room for the ambiguities found in \cite{Russo} depending on whether the conformal condition is imposed before or after worldsheet integration.

Let us now take the heavy string states to be magnon solitons with two angular momenta. In this case the structure constant in the three-point function becomes
\be
{\cal C}_3^{\hbox{\tiny{(primary)}}} = c_{\Delta}^{\hbox{\tiny{(primary)}}}  \big( (\kappa^2 - \cos^2 \beta) 
2I{(\Delta_p,j+2)} - (\kappa^2 - 1)  I{(\Delta_p,j)} \big) \left( \sin \frac {p}{2} \right)^j \ .
\ee
The conformal condition is not a global factor now and thus evaluating the integrals and imposing afterwards the conformal constraint we find a non-vanishing result,
\be
{\cal C}_3^{\hbox{\tiny{(primary)}}} = 2 \pi c_{\Delta}^{\hbox{\tiny{(primary)}}} \, \hbox{sech} \, \alpha \sin^2 \beta \sec \beta
\frac { \Gamma(j/2) \Gamma(1+j/2)^2}{\Gamma(1/2+j/2) \Gamma(3/2+j/2)}
\left( \sin \frac {p}{2} \right)^j \ .
\ee
We note however that the structure of the integrals does not favor any ambiguity in the evaluation of the chiral primary correlator when the heavy states are 
giant magnon solitons.

%%%%%%%%%%%%%%%%%%%%%%%%%%%%%%%%%%%%%%%%%%%%%%%%%%%%%%%%%%%%%%%%%%%%%%%%%%%%%%%%%%%%%%
%%%%%%%%%%%%%%%%%%%%%%%%%%%%%%%%%%%%%%%%%%%%%%%%%%%%%%%%%%%%%%%%%%%%%%%%%%%%%%%%%%%%%%

\subsection{Leading Regge trajectory operator}

Let us now analyze the case of a three-point correlation function where the light vertex corresponds to the insertion of an operator on the leading Regge trajectory. 
In principle we could consider vertex operators representing either string states with spin in $AdS_5$ or with angular momentum in $S^5$. However in the background 
of the giant magnon solitons that we are analyzing in this note a non-trivial contribution is obtained only from 
operators representing string states on the leading Regge trajectory with 
angular momentum $j$ in~$S^5$. The corresponding vertex is \cite{twopointTseytlin,BT}
\be
V_{j}^{\hbox{\tiny{(Regge)}}} =  (Y_+)^{-\Delta_j} (\partial X_x \bar{\partial} X_x)^{j/2} \ , \label{VRegge}
\ee
where the scaling dimension is now $\Delta_{j}=\sqrt{2(j-2)} \lambda^{1/4}$. 
In general the above operator can mix with additional bosonic terms arising from diagonalization of the anomalous dimension operator for the string sigma model,
\footnote{See reference \cite{twopointTseytlin} 
for a more detailed discussion on the structure of the general vertex operator on the leading Regge trajectory and the mixing under one-loop renormalization 
of bosonic operators on the string sigma model on $S^5$.} 
\be
(X_x)^{2p+2q} (\partial X_x)^{j/2-2p} (\bar{\partial}X_x)^{j/2-2q} (\partial X_K \partial X_K)^{p} (\bar{\partial} X_L \bar{\partial} X_L)^{q} \ ,
\label{mixing}
\ee
where $p,q=0,\ldots,j/4$. However, in this section we will only intend to get a qualitative picture of the correlators, and thus for simplicity 
we will ignore the contribution from these additional terms. We will thus proceed first to evaluate the vertex (\ref{VRegge}) in the background of the single-charge 
giant magnon. From solution (\ref{giantmagnon}) we get
\be
(\partial X_x \bar{\partial} X_x)^{j/2} = \left( \hbox{sech} (u_{e}) \tanh (u_{e}) \sin \frac {p}{2} \right)^{j} \ ,
\ee
and the coefficient of the three-point correlation function is then
\be
{\cal C}_{3}^{\hbox{\tiny{(Regge)}}} =  c_{\Delta_j}^{\hbox{\tiny{(Regge)}}} \left( \sin \frac {p}{2} \right)^j
\int_{-\infty}^{\infty} \!\! d \tau_e \int_{-\infty}^{\infty} \!\! d \sigma \left( \hbox{sech} (\kappa \tau_e) \right)^{\Delta_j} \! \left( \hbox{sech} (u_e) \tanh (u_e) \right)^j   .
\label{CRegge}
\ee
As in the previous correlators the integrations over $\sigma$ and $\tau$ factorize. The integral over $\tau_e$ is again given by (\ref{Itau}), while the integral 
over $\sigma$ can be computed using that
\be
\int \!\! dx \sinh^b(a x) \, \hbox{sech}^{2b} (a x) = \frac {i^b \left( \tanh (ax) \right)^{2b-1}}{a(1-2b)} 
{}_2F_1 \left( \! \frac {1}{2} - b, 1 - \frac {b}{2}, \frac {3}{2} - b ; \coth^2(ax) \! \right) \, .
\ee
Evaluating the integrals the normalized structure constant (\ref{CRegge}) becomes 
\be
{\cal C}_{3}^{\hbox{\tiny{(Regge)}}} =  c_{\Delta_j}^{\hbox{\tiny{(Regge)}}} \kappa^{j-2} \, C_R (j) \left( \sin \frac {p}{2} \right)^{j+1} \ ,
\ee
where we have taken into account the conformal constraint, and we have defined
\be
C_R (j) = - 
\frac {i^j 2^{-j}(2j+1) \Gamma(-1/2-j) \Gamma(j) \Gamma(\Delta_j/2)}{\Gamma(1/2+\Delta_j/2)} \cos \left( \frac {j\pi}{2} \right) \ .
\ee
In the case of the two-charge giant magnon soliton an identical computation shows that
\be
{\cal C}_{3}^{\hbox{\tiny{(Regge)}}} =  c_{\Delta_j}^{\hbox{\tiny{(Regge)}}} \kappa^{j-2} \, C_R (j) \, \hbox{sech} \alpha \sec \beta \left( \sin \frac {p}{2} \right)^{j} \ .
\ee
We thus find that the coefficients in the three-point correlator when the light vertex is an operator in the leading Regge trajectory scale as $\kappa^{j-2}$. 
The scaling is preserved when mixing with linear combinations of the operators (\ref{mixing}) is included, because each partial derivative 
in the vertex provides a factor of $\kappa$, while a factor $\kappa^{-2}$ is always obtained upon integration. A similar scaling was also noticed in 
reference \cite{RT}, where the heavy vertices were taken to correspond to folded semiclassical strings with spin in $AdS_5$.

%%%%%%%%%%%%%%%%%%%%%%%%%%%%%%%%%%%%%%%%%%%%%%%%%%%%%%%%%%%%%%%%%%%%%%%%%%%%%%%%%%%%%%
%%%%%%%%%%%%%%%%%%%%%%%%%%%%%%%%%%%%%%%%%%%%%%%%%%%%%%%%%%%%%%%%%%%%%%%%%%%%%%%%%%%%%%

\section{Concluding remarks}

In this note we have studied three-point correlation functions within the strong-coupling regime of the AdS/CFT correspondence. The analysis has been performed 
in the semiclassical approximation where two of the vertex operators in the correlation function are heavy string states carrying conserved charges as large as the 
string tension, while the third vertex is a light operator with fixed conserved charges. We have chosen the heavy vertex operators to correspond to giant magnon 
solitons, with either a single or two different angular momenta in $S^5$. The light vertex has been chosen as the dilaton, the superconformal chiral primary, or an 
operator on the leading Regge trajectory. 
  
An interesting continuation of our analysis could be the obtention of the weak-coupling limit of the coefficients of the three-point functions that we have considered. A possible 
path in this direction could be the general method suggested in \cite{Costa} based on renormalization group arguments after deformation in a conformal field theory. An 
alternative derivation of structure constants in the gauge theory regime is the proposal in \cite{Vieira} based on cutting and gluing integrable spin chains. This approach 
can also be probably employed to find the extension to weak-coupling of the three-point coefficients that we have analyzed. In reference \cite{Russo} the large spin expansion 
of some semiclassical three-point correlators exhibited the same structure as expected on the dual gauge theory side, which allowed to conjecture that the corresponding 
structure constants are protected, and thus remain the same both at strong and weak-coupling. It would also be interesting to find out whether a similar phenomenon holds 
for the giant magnon correlators studied in this note.

We have excluded from the discussion in the main part of the text the case where the light vertex is taken to be a singlet massive scalar operator. 
Singlet massive scalar operators are built from derivatives of the $S^5$ directions and at leading order a possible choice is~\cite{RTvertex,RT}
\be
V^{\hbox{\tiny{(singlet)}}} = (Y_+)^{-\Delta_r} \big( (\partial X_K \partial X_K)  (\bar{\partial} X_L \bar{\partial} X_L) \big)^{r/2} \ , 
\ee
where $r=2, 4, \ldots$, and with $\Delta_r = 2 \sqrt{(r-1)} \lambda^{1/4}$ the corresponding scaling dimension. But the derivative factor is nothing but the classical stress tensor 
for the $S^5$ string sigma model, which is conserved. Therefore a divergent contribution to the coefficient in the three-point function is expected as a consequence of the soliton 
nature of the giant magnon. It would also be of help to uncover and understand this behavior from the gauge theory side.

An additional extension of the work in this note is the study of three-point correlation functions for other choices of giant magnon heavy vertices. More general giant 
magnon solitons with additional angular momenta in $S^5$ or spin in $AdS_5$ where considered in references \cite{KRT}-\cite{dressingII}. The analysis of the 
corresponding correlators could be of help to exhibit general features of the giant magnon three-point structure constants.

%%%%%%%%%%%%%%%%%%%%%%%%%%%%%%%%%%%%%%%%%%%%%%%%%%%%%%%%%%%%%%%%%%%%%%%%%%%%%%%%%%%%%%
%%%%%%%%%%%%%%%%%%%%%%%%%%%%%%%%%%%%%%%%%%%%%%%%%%%%%%%%%%%%%%%%%%%%%%%%%%%%%%%%%%%%%%

\vspace{10mm}
\centerline{\bf Acknowledgments}

This work is supported by MICINN through a Ram\'on y Cajal contract and grant FPA2008-04906, and by 
BSCH-UCM through grant GR58/08-910770. 

%%%%%%%%%%%%%%%%%%%%%%%%%%%%%%%%%%%%%%%%%%%%%%%%%%%%%%%%%%%%%%%%%%%%%%%%%%%%%%%%%%%%%%
%%%%%%%%%%%%%%%%%%%%%%%%%%%%%%%%%%%%%%%%%%%%%%%%%%%%%%%%%%%%%%%%%%%%%%%%%%%%%%%%%%%%%%

\end{document}